\def\ra{\rangle}
\def\la{\langle}
\def\be{\begin{equation}}
\def\ee{\end{equation}}
\def\ba{\begin{array}}
\def\ea{\end{array}}

\def\Rb{{I\!\! R}}

\documentclass[aps,pre,preprint,showpacs,amsmath,amssymb,amsfonts,preprintnumbers]
{revtex4}
\usepackage{graphicx}

\begin{document}

\baselineskip=18pt \setcounter{page}{1} \centerline{\large\bf
Bipartite Bell Inequality and Maximal Violation} \vspace{4ex}
\begin{center}
Ming Li$^{1}$, Shao-Ming Fei$^{2,3}$ and Xianqing Li-Jost$^{3,4}$

\vspace{2ex}

\begin{minipage}{5in}

\small $~^{1}$ {\small College of Mathematics and Computational
Science, China University of Petroleum, 257061 Dongying}

\small $~^{2}$ {\small School of Mathematical Sciences, Capital
Normal University, 100048 Beijing}

{\small $~^{3}$ Max-Planck-Institute for Mathematics in the
Sciences, 04103 Leipzig}

\small $~^{4}$ {\small Department of Mathematics, Hainan Normal
University, 571158 Haikou}
\end{minipage}
\end{center}

\begin{center}
\begin{minipage}{5in}
\vspace{1ex} \centerline{\large Abstract} \vspace{1ex} We present new
bell inequalities for arbitrary dimensional bipartite quantum systems.
The maximal violation of the inequalities is computed.
The Bell inequality is capable of detecting
quantum entanglement of both pure and mixed quantum states more effectively.

\smallskip
PACS numbers: 03.67.-a, 02.20.Hj, 03.65.-w\vfill
\smallskip
\end{minipage}\end{center}
\bigskip

The problem related to the local realism and quantum mechanics was
first highlighted by the paradox of Einstein, Podolsky and Rosen
\cite{1}. Later, Bell proposed a remarkable inequality that
should be obeyed by any local realistic theory \cite{2}. From then
on investigation of Bell theorem for a general quantum system has been
regarded as one of the most important challenges in quantum
mechanics and quantum information science \cite{31,32,33,34,35,36}.

Since Bell's work there have appeared many important further results
such as Clauser-Horne-Shimony-Holt (CHSH) \cite{3} and
Mermin-Ardehali-Belinskii-Klyshko (MABK) inequalities \cite{4}. In
1991 Gisin presented a theorem saying that any pure entangled states
of two spin-1/2 particles (qubits) violate the CHSH inequality
\cite{gisin91}. A more complete and simpler proof of this theorem
had been given for two arbitrary spin-$j$ particles systems
\cite{gisin92}. Soon after a necessary and sufficient condition for
violating the CHSH inequality by an arbitrary mixed
spin-$\frac{1}{2}$ state is presented in \cite{ho340}.  In
\cite{sam92} the authors showed that mixed states can produce
maximal violations in the Bell inequality. Since then there have
been many results on studying the Gisin's theorem for general
bipartite systems and three qubits systems \cite{gisin3q} in terms
of various kinds of Bell inequalities. In the reference \cite{12},
all pure entangled two-qudit states violate an inequality was
studied. A Bell inequality for two $3$-dimensional particles was
also given in \cite{6}. In \cite{jlc} the authors show the maximal
violation of Bell inequalities which corresponds to the maximal
eigenvalue of the Bell operator matrix for two d-dimensional
systems. The maximal violation of the
Collins-Gisin-Linden-Massar-Popescu Inequality \cite{33} is also
studied in \cite{120406}.

In this paper, we present new Bell inequalities for bipartite
quantum states. The maximal violation of the Bell inequalities is
computed analytically for pure states. The inequality and it's
maximal violation are also shown to be valid for all bipartite even
dimensional bipartite mixed states and some odd dimensional
bipartite mixed states. It is shown that some entangled quantum
states that are unrecognizable by using the Bell inequality given in
\cite{gisin92} can be detected by the new Bell inequalities.

We consider bipartite states on $N\times N$ systems. For even $N$, let
$\Gamma_x$, $\Gamma_y$ and $\Gamma_z$ be block-diagonal matrices, in
which each block is an ordinary Pauli matrix, $\sigma_x, \sigma_y$
and $\sigma_z$ repectively, as described in \cite{gisin92} for $\Gamma_x$ and $\Gamma_z$.
When $N$ is odd, we set the elements of the $k$th row and the $k$th column in
$\Gamma_x$, $\Gamma_y$ and $\Gamma_z$ to be zero. The rest elements of
$\Gamma_x$, $\Gamma_y$ and $\Gamma_z$ are
the block-diagonal matrices as for the even $N$ case.
Let $\Pi(k)$ be an $N\times N$ matrix whose only nonvanishing entry is
$(\Pi(k))_{kk}=1$, $k\in 1, 2, \cdots, N$ for odd $N$ and be a null matrix for
even $N$. We define observables
\be\label{1}
A=\vec{a}\cdot\vec{\Gamma}+\Pi(k)=a_x\Gamma_x+a_y\Gamma_y+a_z\Gamma_z+\Pi(k)
\ee
and
\be\label{2}
B=\vec{b}\cdot\vec{\Gamma}+\Pi(k)=b_x\Gamma_x+b_y\Gamma_y+b_z\Gamma_z+\Pi(k),
\ee
where $\vec{a}=(a_x, a_y, a_z)$ and $\vec{b}=(b_x, b_y, b_z)$ are
unit vectors.

We define the Bell operator  as
\be\label{bell}
{\cal B}=A_1\otimes B_1+A_1\otimes B_2+A_2\otimes B_1-A_2\otimes B_2,
\ee
where
\begin{eqnarray*}
A_i&=&\vec{a}_i\cdot\vec{\Gamma}+\Pi(k)=a_i^x\Gamma_x+a_i^y\Gamma_y+a_i^z\Gamma_z +\Pi(k),\\
B_i&=&\vec{b}_j\cdot\vec{\Gamma}+\Pi(k)=b_j^x\Gamma_x+b_j^y\Gamma_y+b_j^z\Gamma_z +\Pi(k).
\end{eqnarray*}

\textbf{\emph{Theorem}}: If there exists local hidden variable model
to describe the system, the inequality
\begin{eqnarray}\label{b22}
|\la {\cal B} \ra|\leq 2
\end{eqnarray} must hold for any $\vec{a}_i$, $\vec{b}_i$, $i=1,2$, and all $k\in 1, 2, \cdots, N$.

The Proof of this theorem is straightforward. Note that for any 3-dimensional
unit vectors $\vec{a}$ and $\vec{b}$, the eigenvalues of $A$ and
$B$ are either $1$ or $-1$. Then as discussed for two-qubit
case, if there exists local hidden variable model to describe the system, one has
\begin{eqnarray*}
|\la {\cal B} \ra|&=&|\la A_1\otimes
B_1+A_1\otimes
B_2+A_2\otimes B_1-A_2\otimes B_2 \ra|\\
&=&|\la A_1\otimes (B_1+B_2)\ra+\la A_2\otimes (B_1-B_2)\ra|\\
&\leq&|\la A_1\ra||\la(B_1+B_2)\ra|+|\la A_2\ra||\la(B_1-B_2)\ra|\leq 2.
\end{eqnarray*}

Now we compute the maximal violation of the Bell inequality.

\textbf{\emph{Proposition 1}}: For any bipartite pure state
$|\psi\ra$ with even $N$, the maximal violation of the Bell
inequality (\ref{b22}) is given by \be\label{maxv} \max\la\psi|{\cal
B}|\psi\ra=2\sqrt{\tau_1+\tau_2}, \ee where $\tau_1$ and $\tau_2$
are the two largest eigenvalues of the matrix $R^TR$, $R$ is the
matrix with entries
$R_{\alpha\beta}=\la\psi|\Gamma_\alpha\otimes\Gamma_\beta|\psi\ra$,
$\alpha,\,\beta=x,y,z$.

{\sf [Proof]} If $N$ is even, we have the maximal violation of the
Bell inequalities (\ref{b22}),
\begin{eqnarray*}
\max\la\psi|{\cal B}|\psi\ra&=&\max_{\vec{a}_1,\vec{a}_2,\vec{b}_1,\vec{b}_2}[\la\psi|\sum\limits_{\alpha=x,y,z}
a_{1}^\alpha\Gamma_\alpha\otimes\sum\limits_{\beta=x,y,z}(b_{1}^\beta+b_{2}^\beta)\Gamma_\beta|\psi\ra\\
&& + \la\psi|\sum\limits_{\alpha=x,y,z}
a_{2}^\alpha\Gamma_\alpha\otimes\sum\limits_{\beta=x,y,z}(b_{1}^\beta-b_{2}^\beta)\Gamma_\beta|\psi\ra]\\
&=&\max_{\vec{a}_1,\vec{a}_2,\vec{b}_1,\vec{b}_2}[\vec{a}_1\cdot R(\vec{b}_1+\vec{b}_2)
+\vec{a}_2\cdot R(\vec{b}_1-\vec{b}_2)]\\
&=&\max_{\vec{b}_1,\vec{b}_2}[||R(\vec{b}_1+\vec{b}_2)||+||R(\vec{b}_1-\vec{b}_2)||]\\
&=&\max_{\theta,\vec{c}\bot\vec{c}^{'}}2[\cos{\theta} ||R\vec{c}||+\sin{\theta}||R\vec{c}^{'}||]\\
&=&\max_{\vec{c}\bot\vec{c}^{'}}2\sqrt{||R\vec{c}||^2+||R\vec{c^{'}}||^2}
=2\sqrt{\tau_1+\tau_2},
\end{eqnarray*}
where $\vec{a}_i=(a_i^x,a_i^y,a_i^z)$,
$\vec{b}_j=(b_j^x,b_j^y,b_j^z)$, $i,j=1,2$.\hfill $\Box$

\textbf{\emph{Proposition 2}}: For any bipartite pure state
$|\Psi\ra$ in the Schmidt bi-orthogonal form,
\be\label{sch}|\Psi\ra=\sum_{i=1}^N
c_i|ii\ra,~~~c_i\in\Rb,~~~\sum_ic_i^2=1
\ee
with odd $N$, the
maximal violation of the Bell inequality (\ref{b22}) is given by
\be\label{maxv2}
\max\la\Psi|{\cal B}|\Psi\ra=
2\sqrt{\tau_1+\tau_2}+2\la\Psi|\Pi(k)\otimes\Pi(k)|\Psi \ra,
\ee
where $\tau_1$ and $\tau_2$ are defined in Proposition 1.

{\sf [Proof]} For odd $N$ any $k\in\{1,2,\cdots,N\}$,
similarly we have
\begin{eqnarray*}
\max\la\Psi|B|\Psi\ra&=&\max_{\vec{a}_1,\vec{a}_2,\vec{b}_1,\vec{b}_2}[\la\Psi|(\sum\limits_{\alpha=x,y,z}
a_{1}^\alpha\Gamma_\alpha+\Pi(k))\otimes(\sum\limits_{\beta=x,y,z}
(b_{1}^\beta+b_{2}^\beta)\Gamma_\beta+2\Pi(k))|\Psi\ra\\
&& + \la\Psi|(\sum\limits_{\alpha=x,y,z}
a_{2}^\alpha\Gamma_\alpha+\Pi(k))\otimes(\sum\limits_{\beta=x,y,z}
(b_{1}^\beta-b_{2}^\beta)\Gamma_\beta)|\Psi\ra]\\
&=&\max_{\vec{a}_1,\vec{a}_2,\vec{b}_1,\vec{b}_2}[\vec{a}_1\cdot R(\vec{b}_1+\vec{b}_2)+\vec{a}_2\cdot
R(\vec{b}_1-\vec{b}_2)]+2\la\Psi|\Pi(k)\otimes\Pi(k)|\Psi\ra\\
&=&\max_{\vec{b}_1,\vec{b}_2}[||R(\vec{b}_1+\vec{b}_2)||+||R(\vec{b}_1-\vec{b}_2)||]
+2\la\Psi|\Pi(k)\otimes\Pi(k)|\Psi\ra\\
&=&\max_{\theta,\vec{c}\bot\vec{c}^{'}}2[\cos{\theta} ||R\vec{c}||+\sin{\theta}
||R\vec{c}^{'}||]+2\la\Psi|\Pi(k)\otimes\Pi(k)|\Psi\ra\\
&=&\max_{\vec{c}\bot\vec{c}^{'}}2\sqrt{||R\vec{c}||^2+||R\vec{c^{'}}||^2}+2\la\Psi|\Pi(k)\otimes\Pi(k)|\Psi\ra\\
&=&2\sqrt{\tau_1+\tau_2}+2\la\Psi|\Pi(k)\otimes\Pi(k)|\Psi\ra.
\end{eqnarray*}
\hfill $\Box$

\emph{Remark:} For even $N$, the formula (\ref{maxv}) is also valid
for any bipartite mixed quantum states $\rho$. One only needs to
redefine $R_{\alpha\beta}= Tr[\rho\Gamma_\alpha\otimes\Gamma_\beta]$
for $\alpha,\beta=x,y,z$. The formula (\ref{maxv2}) doesn't fit for
general quantum states with odd $N$. However, for some quantum mixed
states their maximal violation of the Bell inequality (\ref{b22})
can be still computed by the formula (see the example 2 below).

Moreover the Bell inequality in \cite{gisin92} is a special case of
(\ref{b22}) in the sense that it can be obtained by setting $a_y,
b_y$ in (\ref{1}) and (\ref{2}) to be zero, and $k=N$ in our Bell
operator (\ref{bell}). For $k=N$, the maximal violation of
(\ref{b22}) for an arbitrary bipartite quantum state (\ref{sch}) is
the same as the violation values given in \cite{gisin92}. Which
means that the parameters $a_y$, $b_y$ do not contribute to the
maximal violation in this case. However, even in this case the
formula (\ref{maxv}) and (\ref{maxv2}) have their own advantages. On
one hand, one can compute the maximal violation without choosing
proper Bell operator as is needed in \cite{gisin92}. On the other
hand, for odd $N$, by adjusting $k$ more entangled quantum states
can be detected. In the following we give two examples to show these
properties.

\emph{Example 1:} Consider a $3\times 3$ pure state with Schmidt decomposition
$|\psi\ra=(|11\ra+|33\ra)/\sqrt{2}$.
Using the Bell operator given in \cite{gisin92} one gets
the maximal violation $2$, which fails to detect the entanglement.
Now taking $k=2$ we obtain the maximal violation of our Bell
inequality (\ref{bell}) $2\sqrt{2}$, which means that $|\psi\ra$ is entangled.

Our Bell inequality valids also for all mixed states
with even $N$ and for some mixed states with odd $N$. Therefore
it can be used to detect experimentally the entanglement of mixed states.

\emph{Example 2:} Consider that the maximally entangled state
$|\psi_+\ra=\sum\limits_{i=1}^N\frac{1}{\sqrt{N}}|ii\ra$ mixed with
noise:
\be
\rho(x)=\frac{x}{N^2}I+(1-x)|\psi_+\ra\la\psi_+|.
\ee
For even $N$, the maximal violation of $\rho(x)$ is
$2\sqrt{2}(1-x)$. Therefor, Bell inequality (\ref{b22}) detect
entanglement of $\rho(x)$ for $0\leq x<0.292893$. If $N$ is odd, first
note that for any $k\in\{1,2,\cdots,N\}$ and $\alpha\in\{x, y, z\}$,
$(\Gamma_\alpha)_{kk}=0$. Thus we have
\be\label{ex}
{\rm Tr}[\rho(x)(\Gamma_\alpha\otimes\Pi(k))]={\rm
Tr}[\rho(x)(\Pi(k)\otimes\Gamma_\alpha)]=0.
\ee
Taking into account (\ref{ex}) we have the maximal violation
\begin{eqnarray*}
\max{\rm Tr} [\rho(x)B]&=&\max_{\vec{a}_1,\vec{a}_2,\vec{b}_1,\vec{b}_2}
\left\{{\rm Tr} [\rho(x)(\sum\limits_{\alpha=x,y,z}
a_{1}^\alpha\Gamma_\alpha+\Pi(k))\otimes(\sum\limits_{\beta=x,y,z}
(b_{1}^\beta+b_{2}^\beta)\Gamma_\beta+2\Pi(k))]\right.\\
&&\left.  + {\rm Tr} [\rho(x)(\sum\limits_{\alpha=x,y,z}
a_{2}^\alpha\Gamma_\alpha+\Pi(k))\otimes(\sum\limits_{\beta=x,y,z}
(b_{1}^\beta-b_{2}^\beta)\Gamma_\beta)]\right\}\\
&=&\max_{\vec{a}_1,\vec{a}_2,\vec{b}_1,\vec{b}_2}[\vec{a}_1\cdot R(\vec{b}_1+\vec{b}_2)+\vec{a}_2\cdot
R(\vec{b}_1-\vec{b}_2)]+2{\rm Tr} [\rho(x)\Pi(k)\otimes\Pi(k)]\\
&=&\max_{\vec{c}\bot\vec{c}^{'}}2\sqrt{||R\vec{c}||^2+||R\vec{c^{'}}||^2}+2{\rm Tr} [\rho(x)\Pi(k)\otimes\Pi(k)]\\
&=&2\sqrt{\tau_1+\tau_2}+2{\rm Tr} [\rho(x)\Pi(k)\otimes\Pi(k)],
\end{eqnarray*}
where $R_{\alpha\beta}= Tr[\rho\Gamma_\alpha\otimes\Gamma_\beta]$. For $N=3$, the maximal
violation of $\rho(x)$ is
$\frac{2}{25}(5-4x)+\frac{8\sqrt{2}}{5}(1-x)$. Hence the Bell inequality
(\ref{b22}) can detect then the entanglement of
$\rho(x)$ for $0\leq x<0.2566$ in this case.

\medskip

We have studied bipartite Bell inequality by constructing new Bell
operator which includes the Gisin's Bell inequalities in
\cite{gisin92} as a special case. The maximal violation of these
Bell inequalities for pure states in Schmidt forms has been
obtained. The formulae of maximal violation valid also for all pure
and mixed quantum states in even dimensional bipartite systems and
for some mixed states in odd dimensional bipartite ones. The new
Bell inequality has been shown to be capable of detecting quantum
entanglement more effectively.

\bigskip
\noindent{\bf Acknowledgments}\,
This work is supported by the NSFC 10675086, NSFC 10875081 and KZ200810028013.


\smallskip
\begin{thebibliography}{99}

\bibitem{1} A. Einstein, B. Podolsky, and N. Rosen, Phys. Rev. 47,
777 (1935).

\bibitem{2} J. S. Bell, Physics 1, 195 (1964).

\bibitem{31} N. D. Mermin, Phys. Rev. D 22, 356 (1980); A. Gard and
N. D. Mermin, Found. Phys. 14, 1 (1984).

\bibitem{32} D. Kaszlikowski, P
Gnacinski, M. Zukowski, W. Miklaszewski, and A. Zeilinger, Phys.
Rev. Lett. 85, 4418 (2000).

\bibitem{33} D. Collins, N. Gisin, N. Linden, S.
Massar, and S. Popescu, Phys. Rev. Lett. 88, 040404 (2002).

\bibitem{34} N. D. Mermin, Phys. Rev. Lett. 65, 1838 (1990).

\bibitem{35} M. Ardehali, Phys.
Rev. A 46, 5375 (1992).

\bibitem{36}N. J. Cerf, S. Massar, and S. Pironio, Phys.
Rev. Lett. 89, 080402 (2002).

\bibitem{3} J. F. Clauser, M. A. Horne, A. Shimony, and R. A. Holt,
Phys. Rev. Lett. 23, 880 (1969).

\bibitem{4} N. D. Mermin, Phys. Rev. Lett. 65, 1838 (1990); S. M.
Roy and V. Singh, ibid. 67, 2761 (1991); M. Ardehali, Phys. Rev. A
46, 5375 (1992); A. V. Belinskii and D. N. Klyshko, Phys. Usp. 36,
653 (1993); N. Gisin and H. Bechmann-Pasquinucci, Phys. Lett. A 246,
1 (1998).

\bibitem{gisin91} N. Gisin, Phys. Lett. A, 154, 201(1992).


\bibitem{gisin92} N. Gisin and A. Peres, Phys. Lett. A, 162, 15(1992).


\bibitem{ho340} R. Horodecki, P. Horodecki and M. Horodecki, Phys. Lett.
A, 200, 340(1995).

\bibitem{sam92} S. L. Braunstein, A. Mann and M. Revzen, Phys. Rev.
Lett. 68, 3259(1992).


\bibitem{gisin3q}  J. L. Chen, C. F. Wu, L. C. Kwek, and C. H. Oh, Phys. Rev. Lett. 93, 140407(2004).

\bibitem{12} J. L. Chen, D. L. Deng, and M. G. Hu, Phys. Rev. A, 77, 060306(R) (2008).

\bibitem{6} D. Kaszlikowski, L. C. Kwek, J. L. Chen, M. Zukowski,
and C. H. Oh, Phys. Rev. A 65, 032118 (2002).

\bibitem{jlc}  J. L. Chen, C. F. Wu, L. C. Kwek, C. H. Oh£¬and M. L. Ge, Phys. Rev. A, 74, 032106
(2006).


\bibitem{120406} S. Zohren, R. D. Gill, Phys. Rev. Lett. 100, 120406 (2008).
\end{thebibliography}
\end{document}